\documentclass[twocolumn]{aastex63}
\usepackage[utf8]{inputenc}
\usepackage{graphicx}
\usepackage[T1]{fontenc}
\usepackage{calc}
\usepackage{enumitem}
\usepackage{amsmath}
\usepackage{amssymb}
\usepackage{float}
\usepackage[english]{babel}
\usepackage{array}
\usepackage{placeins}
\usepackage{multirow}
\usepackage{xcolor}
\usepackage{textcomp}
\usepackage{gensymb}

\def\mx{MAXI~J1820$+$070}
\def\xte{XTE~J1550$-$564}
\def\hh{H1743$-$322}

\begin{document}

\title{Relativistic X-ray jets from the black hole X-ray binary \mx{}}

\author[0000-0001-9075-1489]{Mathilde Espinasse}
\affiliation{AIM, CEA, CNRS, Université Paris-Saclay, Université Paris Diderot, Université de Paris, F-91191 Gif-sur-Yvette, France}

\author[0000-0001-5538-5831]{Stéphane Corbel}
\affiliation{AIM, CEA, CNRS, Université Paris-Saclay, Université Paris Diderot, Université de Paris, F-91191 Gif-sur-Yvette, France}
\affiliation{Station de Radioastronomie de Nançay, Observatoire de Paris, PSL Research University, CNRS, Univ. Orléans, 18330 Nançay, France}

\author[0000-0002-3638-0637]{Philip Kaaret}
\affiliation{Department of Physics and Astronomy, University of Iowa, Iowa City, IA 52242, USA}

\author{Evangelia Tremou}
\affiliation{AIM, CEA, CNRS, Université Paris-Saclay, Université Paris Diderot, Université de Paris, F-91191 Gif-sur-Yvette, France}
\affiliation{LESIA, Observatoire de Paris, CNRS, PSL, SU/UPD, Meudon, France}

\author{Giulia Migliori}
\affiliation{INAF, Istituto di Radio Astronomia di Bologna, Via P. Gobetti 101, I-40129 Bologna, Italy}

\author{Richard M. Plotkin}
\affiliation{Department of Physics, University of Nevada, Reno, Nevada, 89557, USA}

\author{Joe Bright}
\affiliation{Astrophysics, Department of Physics, University of Oxford, Keble Road, Oxford, OX1 3RH, UK}

\author{John Tomsick}
\affiliation{Space Sciences Laboratory, 7 Gauss Way, University of California, Berkeley, CA 94720-7450, USA}

\author{Anastasios Tzioumis}
\affiliation{Australia Telescope National Facility, CSIRO, PO Box 76, Epping, New South Wales 1710, Australia}

\author{Rob Fender}
\affiliation{Astrophysics, Department of Physics, University of Oxford, Keble Road, Oxford, OX1 3RH, UK}
\affiliation{Department of Astronomy, University of Cape Town, Private Bag X3, Rondenbosch, 7701, South Africa}

\author[0000-0001-9647-2886]{Jerome A. Orosz}
\affiliation{Department of Astronomy, San Diego State University, 5500 Campanile Drive, San Diego, CA 92182-1221, USA}

\author{Elena Gallo}
\affiliation{Department of Astronomy, University of Michigan, 1085 S University, Ann Arbor, MI 48109, USA}

\author{Jeroen Homan}
\affiliation{Eureka Scientific, Inc., 2452 Delmer Street, Oakland, CA 94602, USA}
\affiliation{SRON Netherlands Institute for Space Research, Sorbonnelaan 2, NL-3584 CA Utrecht, the Netherlands}

\author{Peter G. Jonker}
\affiliation{Department of Astrophysics/IMAPP, Radboud University, PO Box 9010, NL-6500 GL Nijmegen, the Netherlands}
\affiliation{SRON Netherlands Institute for Space Research, Sorbonnelaan 2, NL-3584 CA Utrecht, the Netherlands}

\author{James C. A. Miller-Jones}
\affiliation{International Centre for Radio Astronomy Research, Curtin University, GPO Box U1987, Perth, WA 6845, Australia}

\author{David M. Russell}
\affiliation{Center for Astro, Particle and Planetary Physics, New York University Abu Dhabi, PO Box 129188, Abu Dhabi, UAE}

\author{Sara Motta}
\affiliation{Astrophysics, Department of Physics, University of Oxford, Keble Road, Oxford, OX1 3RH, UK}

\begin{abstract}
The black hole \mx{} was discovered during its 2018 outburst and was extensively monitored across the electromagnetic spectrum. Following the detection of relativistic radio jets, we obtained four \textit{Chandra} X-ray observations taken between 2018 November and 2019 May, along with radio observations conducted with the VLA and MeerKAT arrays.
We report the discovery of X-ray sources associated with the radio jets moving at relativistic velocities with a possible deceleration at late times. The broadband spectra of the jets are consistent with synchrotron radiation from particles accelerated up to very high energies ($>$ 10 TeV) by shocks produced by the jets interacting with the interstellar medium. The minimal internal energy estimated from the X-ray observations for the jets is $\sim 10^{41}$ erg, significantly larger than the energy calculated from the radio flare alone, suggesting most of the energy is possibly not radiated at small scales but released through late-time interactions. \\
\end{abstract}

\section{Introduction} 

Jets and outflows are observed in a diverse range of accreting systems such as young stellar objects, Galactic X-ray binaries and active galactic nuclei (AGN). The formation of jets, their propagation and their association with accretion processes are still largely unclear. However, their feedback on their immediate environment is now starting to be quantified, as their interaction with the interstellar medium can be observed using high spatial resolution images of X-ray binaries \citep{Corbel2002, Migliori2017}. 
Large-scale Galactic jets with apparent superluminal motion were originally detected in GRS 1915+105 by \citet{Mirabel1994}. Such jets originate in discrete ejecta launched during state transitions \citep{Corbel2004, Fender2004}. The associated radio emission is characteristic of evolving synchrotron blobs \citep{vanderLaan1966} whose fate was unclear until the detection of their reactivation at high energy when they interact with the interstellar medium, e.g. \xte{} \citep{Corbel2002,Tomsick2003, Kaaret2003, Migliori2017} and \hh{} \citep{Corbel2005}.

\mx{}, first known as ASASSN$-$18ey, is a black hole X-ray binary \citep{Tucker2018,Torres2019} originally discovered in the optical band on 2018 March 6 by the All-Sky Automated Survey for Supernovae ASAS-SN \citep{Shappee2014, Kochanek2017} and in X-rays on 2018 March 11 \citep{Kawamuro2018, Denisenko2018} by the Monitor of All-sky X-ray Image MAXI on board the International Space Station \citep{Matsuoka2009}. Its distance is constrained to 2.96 $\pm$ 0.33 kpc by radio parallax measurements \citep{Atri2020}.
Its 2018 and 2019 outbursts were densely monitored in radio, revealing the ejection of long-lived discrete relativistic jets \citep{Bright2020}. The discovery of these jets in the radio wavelengths motivated the search for X-ray counterparts.

%%%%%%%%%%%%%%%%%%%%%%%%%%%%%%%

\begin{figure*}[ht]
    \centering
    \includegraphics[trim={0 5cm 0 5cm}, clip, width=\linewidth]{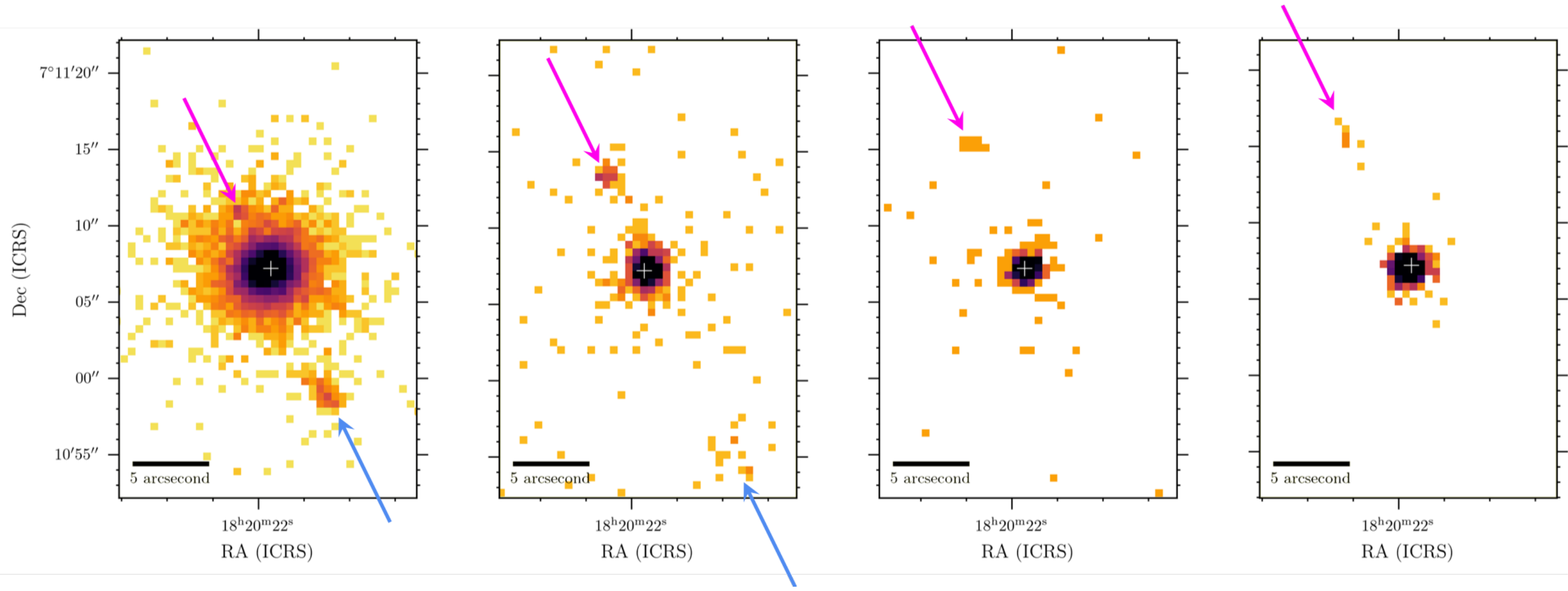}
    \caption{Images obtained from \textit{Chandra} observations of \mx{} in the 0.3 -- 8 keV band. The observations are in chronological order : 2018 November, 2019 February, May and June. The color scale is logarithmic and different for every image. The crosses indicate the VLBI position of \mx{} \citep{Atri2020}. The arrows highlight the position of the north (pink) and south (blue) detected sources. The significances are, for the north and south jets, 46 and 43 (109 and 190 photons) in November and 16 and 4.2 (35 and 15 photons) in February ; and for the north jet, 3.5 (6 photons) in May and 4.9 (12 photons) in June.}
    \label{carteX}
\end{figure*}

\section{Observations}

\subsection{X-ray observations with \textit{Chandra} } 

Following the detection of extended radio jets in \mx{}, we triggered our \textit{Chandra} program (PI: S. Corbel) to search for associated extended X-ray emission from the jets. The observations were performed with the Advanced CCD Imaging Spectrometer (ACIS, \citet{Townsley2000}) on 2018 November 13 (30 ks; ObsId 20207) and 2019 February 4 and 5 (2 $\times $ 20 ks; ObsId 20208 and 22080). In addition, we also used two complementary ACIS-S observations (PI: E. Gallo) scheduled during the outburst decay of \mx{} on 2019 May 24 (12 ks; ObsId 21203) and 2019 June 11 (65 ks; ObsId 21205). The remaining \textit{Chandra} observations of the field did not reveal any X-ray sources besides \mx{} ; all observations are summarized in Table \ref{tableau1}. All observations were performed in sub-array mode to minimize pile-up and we restricted the analysis to the S3 chip, which provides the best low-energy response. 

The X-ray data analysis was performed using the \textit{Chandra} Interactive Analysis of Observation (CIAO) software 4.11 \citep{Fruscione2006}, with the calibration files CALDB version 4.8.4.1. The $\texttt{chandra\_repro}$ script was run to reprocess the observations. The 2018 November observation was processed to remove the ACIS readout streak and the two observations taken in 2019 February were merged. Then, all observations were filtered to keep only the events in the energy range 0.3 -- 8 keV. The \texttt{fluximage} script was used to create the X-ray images, keeping the bin size to 1 (1 pixel = 0.492\arcsec).

\subsection{Radio observations with VLA and MeerKAT}

\mx{} was also observed in the radio wavelengths (see Table \ref{tableau1}). We use two observations performed with the Karl G. Jansky Very Large Array (VLA) \citep{Perley2011}. The observations were almost simultaneous with the first and second \textit{Chandra} and MeerKAT observations, with an on-target duration of 30 minutes on 2018 November 9 and 38 minutes on 2019 February 2. The array was in D configuration (synthesized beam of 12\arcsec) during the November observation and in C configuration (synthesized beam of 3.5\arcsec) during the February observation. Both observations were carried out in the C band, which is centered at 6 GHz. The total VLA bandwidth is 4.096 GHz, divided into 32 spectral windows of 128 MHz, each divided again into 64 channels of 2 MHz. \\

MeerKAT \citep{Jonas2016, Camilo2018, Mauch2020} is an array made of 64 13.5m antennas located in Northern Cape, South Africa. Its spatial resolution is around 5\arcsec. Its bandwidth is divided into 4096 channels of 209~kHz, making a total bandwidth of 856 MHz centered at 1.284 GHz (L band). \mx{} was observed regularly during its outbursts as part of the ThunderKAT Large Survey Project \citep{Fender2017}, and we focus here on the observations taken on 2018 November 13 lasting 45 minutes and on 2019 February 1 lasting 15 minutes. Radio flux densities, $S_{\nu}$, and radio spectral indices, $\alpha_r$, defined as $S_{\nu} \propto \nu ^{\alpha_r}$,  are reported in Table \ref{tableau2}. See \citet{Bright2020} for details on the radio monitoring and results. \\

The Common Astronomy Software Applications package (CASA) version 5.1.1-5 was used for all the radio data reduction \citep{McMullin2007}. 
The data were calibrated using the flux calibrators 3C286 for the VLA and PKS B1934-638 for MeerKAT, and the phase calibrators J1824+1044 for the VLA and J1733-1304 for MeerKAT. 
Images were produced from calibrated data using the algorithm \texttt{CLEAN} \citep{Hogbom1974} within CASA. We chose cells of 1.5\arcsec~ for MeerKAT images. The VLA data was divided into 2 sub-bands of 16 spectral windows each, the first sub-band centered on 5 GHz and the second on 7 GHz, approximately. The sub-bands were imaged separately to reduce artefacts, with cells of 2.5 and 1.6\arcsec~ respectively in D configuration, and 0.7\arcsec~ and 0.5\arcsec~ in C configuration. A robust weighting \citep{Briggs1995} of $-0.7$ was used for all images.

%%%%%%%%%%%%%%%%%%%%%%%%%%%%%%%
\section{Results}

\subsection{Source detection}

The CIAO tool \texttt{wavdetect} was used to identify the X-ray sources in the \textit{Chandra} observations. In a 30\arcsec~ radius around the position of \mx{}, three aligned X-rays sources were detected in the 2018 November and 2019 February images. One of them was consistent with the location of \mx{} and the other two moved between November and February. In 2019 May and June, only two sources were detected, one of them consistent with the position of \mx{} and the other to the north with a larger displacement compared to the previous observations.
The angle between the axis of the aligned sources and the north is $25.1 \pm 1.4\degree$.

The images obtained are displayed in Figure \ref{carteX}.
The angular separations between the core source and the other detected sources are listed in Table \ref{tableau2}.  In the following, we refer to the moving sources as the north jet and the south jet, based on their location with respect to \mx{}. 

The source detection process was similar for all radio images. Due to the lower spatial resolution in the radio maps compared with \textit{Chandra}, we used the \textit{Chandra} locations to constrain the components of the radio maps. 
The \texttt{imfit} CASA task was used to perform 2D Gaussian fits. Point sources (2D Gaussians of the size of the beam) were first fitted on all the fixed core positions coming from the \textit{Chandra} data. Then, the residual images were examined and point sources were fitted on the fixed \textit{Chandra} jet positions. The radio fluxes obtained through these fits are presented in Table \ref{tableau2}. 

\subsection{Spectral analysis}

X-ray source and background spectra were extracted for the three detected objects using the \texttt{specextract} script. For all sources in all observations, except the north jet in November, a circular background was chosen from a source-free area of the chip. 

\begin{figure}[ht]%*
  \centering
  \includegraphics[trim = 1.8cm 1cm 1.8cm 1cm, clip, width=\linewidth]{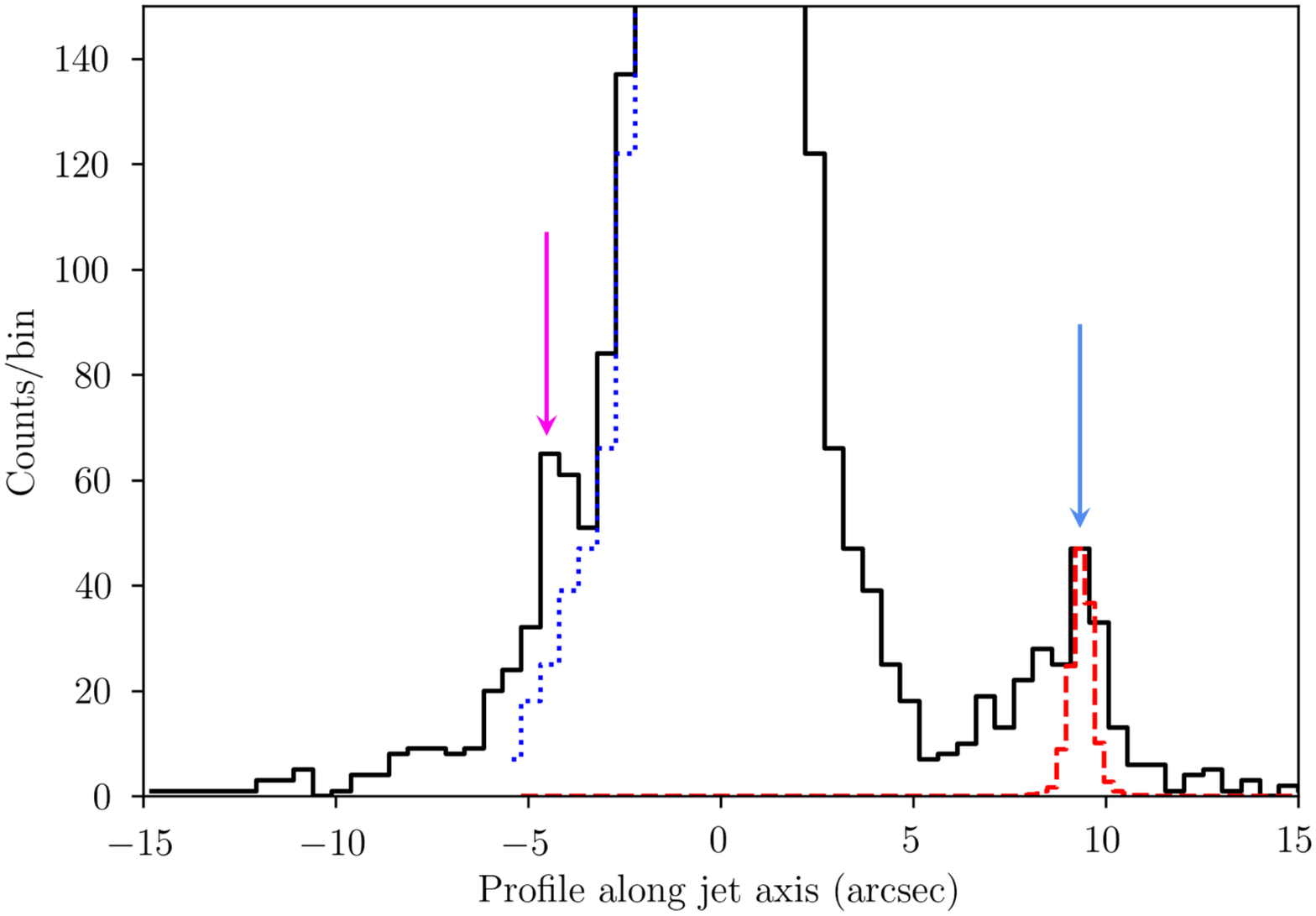}
  \\
  \includegraphics[trim = 7.6cm 5cm 7.8cm 5cm, clip, width=\linewidth]{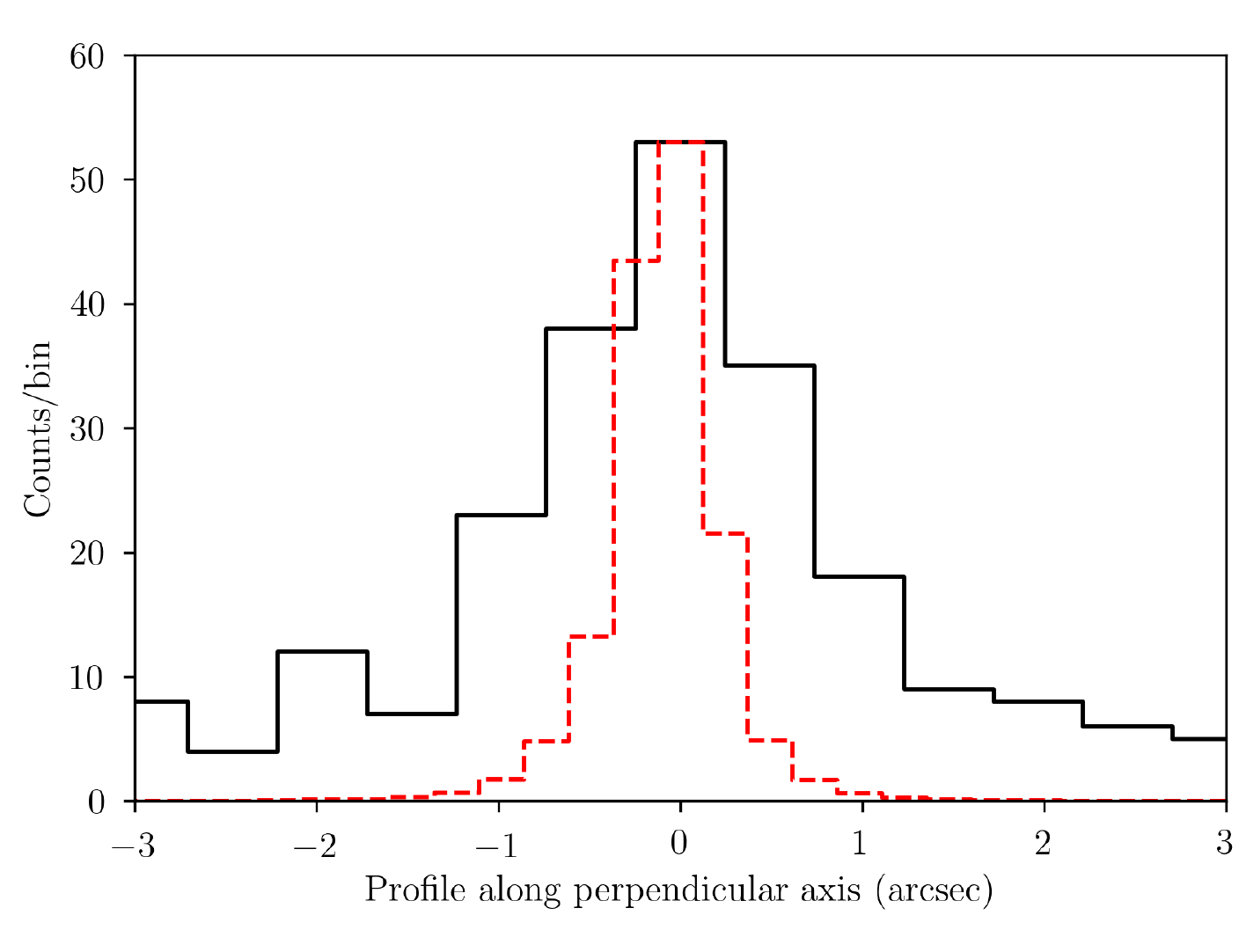}
  \caption{\textit{Top} : Profile taken along the jet axis in 2018 November. The figure has a linear scale, and the y-axis is adapted to make the peaks of the jets visible. The dotted blue line is the mirrored profile of the other side of the PSF, to visually highlight the north jet.
  \textit{Bottom}: Profile perpendicular to the jet axis for the south jet in 2018 November. On both panels, the dashed red line is the rescaled profile of PSF simulated with MARX, with a sub-pixel resolution of half an ACIS pixel.}
  \label{profil}
\end{figure}

As the profile of the north jet in the November observation (Figure \ref{profil}) revealed an overlap with the wings of the central black hole, its background spectrum was extracted from a partial annulus around the black hole (inner radius of 3.2\arcsec~ and outer radius of 5.2\arcsec, subtracting the elliptic region of the north jet). 
The X-ray spectral analysis was then performed using \textit{XSPEC} \citep{Arnaud1996} and \textit{Sherpa}, the CIAO modeling and fitting application developed by the \textit{Chandra X-ray Center} \citep{Freeman2001}.

The spectra extracted from the north and south jets were fitted with an absorbed power-law model with photon index $\Gamma$ (\texttt{tbabs * powerlaw}), using the abundances from \citet{Wilms2000}. The hydrogen column density value of $2.16 ^{+0.73} _{-0.65} \times 10^{21}$ cm$^{-2}$ was obtained by fitting the spectra of \mx{} in 2019 February using statistic \texttt{cstat}, as it did not suffer from the photon pile-up effect that is seen for bright sources. We afterwards froze the hydrogen column density to the best fit value.
The 0.3 -- 8 keV unabsorbed flux and the photon index $\Gamma$ obtained are reported in Table \ref{tableau2}. \\

\begin{deluxetable*}{c c c c c c c}
\tablewidth{\textwidth}
\tablecolumns{10}
\tablecaption{\textit{Chandra}, MeerKAT and VLA observations of \mx{}.}
\label{tableau1}

\tablehead{
  \vspace{-0.65cm} & & \colhead{\textit{Chandra} date} & & \colhead{Frame time} & \colhead{MeerKAT date} & \colhead{VLA date} \\
  \colhead{Obs no.} & \colhead{\textit{Chandra} ObsId} & & \colhead{Subarray} & \vspace{-0.2cm} & & \\
  & & \colhead{(dd/mm/yy)} & & (s) & \colhead{(dd/mm/yy)} & \colhead{(dd/mm/yy)} }
\startdata
     1 & 20207 & \begin{tabular}{@{}c@{}}13--14/11/18 \\ (MJD 58436.0)\end{tabular} & 1/4 & 0.8 & \begin{tabular}{@{}c@{}}13/11/18 \\ (MJD 58435.7)\end{tabular} & \begin{tabular}{@{}c@{}}09/11/18 \\ (MJD 58431.9)\end{tabular} \\
     \hline
     \multirow{4}{*}{\vspace{0.4cm} 2} & 20208 & \begin{tabular}{@{}c@{}}04/02/19 \\ (MJD 58518.7)\end{tabular} & 1/4 & 0.8 & \multirow{4}{*}{\vspace{0.5cm} \begin{tabular}{@{}c@{}}01/02/19 \\ (MJD 58515.2)\end{tabular}} & \multirow{4}{*}{\vspace{0.5cm} \begin{tabular}{@{}c@{}}02/02/19 \\ (MJD 58516.7)\end{tabular}} \\ 
      & 22080 & \begin{tabular}{@{}c@{}}05/02/19 \\ (MJD 58519.2)\end{tabular} & 1/4 & 0.8 & & \\ 
     \hline
     3 & 21203 & \begin{tabular}{@{}c@{}}24/05/19 \\ (MJD 58627.5)\end{tabular} & 1/8 & 0.7 & \nodata & \nodata \\
     \hline
     4 & 21204 & \begin{tabular}{@{}c@{}}02--03/06/19 \\ (MJD 58636.9)\end{tabular} & 1/8 & 0.6 & \nodata & \nodata \\
     \hline
     5 & 21205 & \begin{tabular}{@{}c@{}}11/06/19 \\ (MJD 58645.5)\end{tabular} & 1/8 & 0.6 & \nodata & \nodata \\
\enddata

\end{deluxetable*} %

\begin{deluxetable*}{c c c c c c c c c}
\tablewidth{\textwidth}
\tablecolumns{10}
\tablecaption{Jet detections.}
\label{tableau2}

\tablehead{
  \vspace{-0.65cm} & & \colhead{Separation} & & \colhead{0.3 -- 8 keV flux} & \colhead{1.3 GHz flux} & \colhead{5 GHz flux} & \colhead{7 GHz flux} & \\
   \colhead{Observation} & \colhead{Source} & \vspace{-0.2cm} & \colhead{$\Gamma$} & & & & & \colhead{$\alpha_r$} \\
  & & \colhead{(\arcsec)} &  & \colhead{($10^{-14}$ erg cm$^{-2}$ s$^{-1}$)} & \colhead{(mJy)} & \colhead{(mJy)} & \colhead{(mJy)} & }
\renewcommand{\arraystretch}{1.3}
%\colhead{($10^{-14}$ erg cm$^{-2}$ s$^{-1}$)}
\startdata
     \multirow{2}{*}{1} & South jet & 8.81 $\pm$ 0.06 & $1.95 ^{+0.26} _{-0.25}$ & $10.5 ^{+3.3} _{-2.7}$ & $0.24 \pm 0.03$ & $0.09 \pm 0.02$ & $0.09 \pm 0.02$ & $-0.62 \pm 0.11$ \\
      & North jet & 4.27 $\pm$ 0.04 & $1.52 ^{+0.27} _{-0.28}$ & $7.61 ^{+1.73} _{-1.77}$ & $0.49 \pm 0.03$ & $0.23 \pm 0.02$ & $0.24 \pm 0.02$ & $-0.45 \pm 0.05$ \\
     \hline
     \multirow{2}{*}{2} &  South jet & 12.82 $\pm$ 0.22 & $2.60 ^{+1.19} _{-1.07}$ & $0.73 ^{+0.60} _{-0.60}$ & $<0.1$\tablenotemark{a} & $<0.03$\tablenotemark{a} & $<0.03$\tablenotemark{a} & \nodata \\
      & North jet & 6.57 $\pm$ 0.09 & $1.70 ^{+0.60} _{-0.59}$ & $1.48 ^{+1.28} _{-0.78}$ & $0.17 \pm 0.03$ & $0.05 \pm 0.01$ & $0.04 \pm 0.01$ & $-0.87 \pm 0.15$ \\
     \hline
     3 & North jet & 9.02 $\pm$ 0.12 & $1.6$\tablenotemark{b} & $1.00 ^{+0.64} _{-0.72}$ & \nodata & \nodata & \nodata & \nodata \\
     \hline
     5 & North jet & 9.85 $\pm$ 0.16 & $1.6$\tablenotemark{b} & $0.28 ^{+0.14} _{-0.15}$ & \nodata & \nodata & \nodata & \nodata \\
\enddata

\tablecomments{Separation is the angular separation with the main source in arcseconds. The unabsorbed X-ray flux between 0.3 keV and 8 keV is in units of $10^{-14}$ erg cm$^{-2}$ s$^{-1}$. $\alpha_r$ is the radio spectral index. $\Gamma$ is the X-ray photon index.} \tablenotetext{a}{No detection, $3 \sigma$ upper limit.} \tablenotetext{b}{$\Gamma$ fixed to 1.6 due to the low number of photons.}

\end{deluxetable*} %

%%%%%%%%%%%%%%%%%%%% FIN TEST TABLEAU %%%%%%%%%%%%%%%%%%%%%%%

\subsection{Morphology}

As the angular resolution of the X-ray images is higher than that of the radio images, we studied the morphology of the north and south jets using solely the \textit{Chandra} data.  We extracted from each \textit{Chandra} image the profile along the axis formed by the jets and summed over 4" in width. We used the profile of \mx{} as an estimate of the point spread function (PSF) of the \textit{Chandra} instrument for all observations except for 2018 November which suffered from strong pile-up.  We thus used MARX version 5.4.0 \citep{Davis2012} to simulate the \textit{Chandra} PSF without pile-up for that specific observation.  The PSF profile is then rescaled and plotted over the jets profile to estimate the jet extension.

As an example, the profile obtained from 2018 November, with the PSF overlaid on the south jet, is displayed in Figure \ref{profil} (top panel). 
A Kolmogorov-Smirnov (KS) test was performed comparing the jet profiles against the PSF profiles to determine whether the jets are resolved, the null hypothesis being that the two samples are drawn from the same distribution.
The south jet in 2019 February is too faint with photons widely dispersed to make the test conclusive, though it appears resolved in the image.
According to the results of the other KS tests, only the south jet in 2018 November has a significantly different distribution from the PSF, with a p-value of $7.86 \times 10^{-3}$, which indicates that it is resolved at the 95\% confidence level.

As the south jet is resolved along the axis of the jets in 2018 November, we also compute its profile perpendicularly to that axis (Figure  \ref{profil} bottom panel). The KS test of the jet profile against the PSF profile along a perpendicular axis allows us to reject the null hypothesis that both samples come from the same distribution, with a p-value of $5.23 \times 10^{-5}$ at the 95\% confidence level. 

The south jet is thus resolved in the November \textit{Chandra} observation, with a size of $4.9"$ in length and $2.6"$ perpendicularly to its axis (see section \ref{4.2} for further discussion about the resolved jet). The error on these dimensions can be estimated as the bin width, i.e. $0.5"$. Due to the modeling of this jet as a truncated cone, these dimensions yield an opening angle of the jet of $6.7 \pm 1.4$\degree. \\

%%%%%%%%%%%%%%%%%%%%%%%%%%%%%%%
\section{Discussion}

New \textit{Chandra} observations of \mx{} conducted during the decays of its 2018 and 2019 outbursts led to the detection of two new and variable X-ray sources moving away from the central black hole.  
These sources are consistent with the positions of the radio jets that have been observed by \citet{Bright2020}. In the \textit{Chandra} observations in November, the south jet is resolved both parallel and perpendicular to the jet axis. It is likely we are witnessing the interaction of the jets of \mx{} with the interstellar medium (ISM), similar to that observed from \xte{} \citep{Corbel2002, Kaaret2003, Tomsick2003, Migliori2017} and \hh{} \citep{Corbel2005}.

\subsection{Motion of the jets}

\begin{figure}[ht]
    \centering
    \includegraphics[trim={0cm 7.5cm 0cm 7cm}, clip,width=\linewidth]{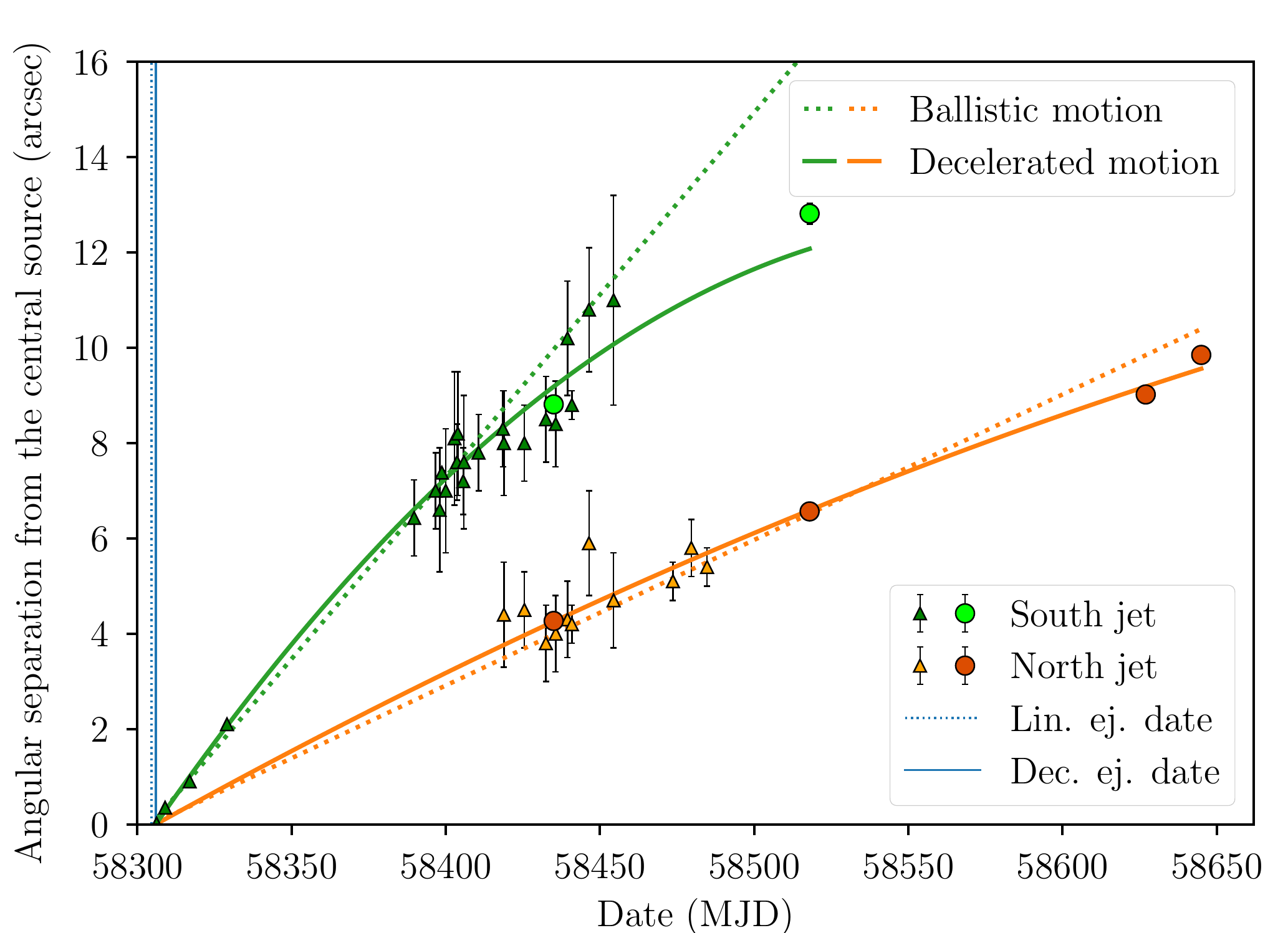}
 \caption{Apparent motion of the two jets. The triangles are the radio data from \citet{Bright2020} and the circles are our X-ray data. The dashed line is the fit for the linear motion and the solid line is the fit for the decelerated motion. The blue lines represent the likely ejection date for the jets in both models. }
 \label{decel}
\end{figure}

The angular separations obtained from the \textit{Chandra} images and presented in Table \ref{tableau2} imply that the two jets are moving. 
We study their apparent motions from the \textit{Chandra} data and using the angular separations obtained in radio by \citet{Bright2020}. Our data extend the time coverage with much later observations.
For both jets, the angular separation versus time is plotted on Figure \ref{decel}. 

We model the motion of the jets as having either constant velocity ($y(t) = v_0 \times (t-t_0)$) or constant deceleration ($y(t) = \frac{1}{2} \dot{v}_0 \times (t- t_0)^2 + v_0 \times (t-t_0)$ with $\dot{v}_0 < 0$). The first fit performed (dashed line in Figure \ref{decel}) is a linear fit, assuming a common ejection date and a ballistic motion for both jets. The joint fits yields an ejection date of MJD $58304.59 \pm 0.08$. Furthermore, we fit the data with a constant deceleration model assuming also a common ejection date for both jets (but not necessarily the same as the launch date in the linear model), and the joint fits give MJD $58305.97 \pm 0.07$ as ejection date in the decelerated model.

The ejection date obtained with the decelerated model is consistent with the one (MJD $58306.03 \pm 0.02$) that \citet{Bright2020} measured without the \textit{Chandra} data, and which occurred during the hard to soft state transition period according to \citet{Shidatsu2019}. On the other hand, the ejection date obtained using the linear model appears one day too early according to the flare observed by \citet{Bright2019}.
This is due to the fact that the \textit{Chandra} position of the jets at the late time disfavours the ballistic model and therefore strengthens the model with jets interacting with the ISM. 

The proper motions obtained with the ballistic fit are $v_{north} = 30.5 \pm 0.2$ mas d$^{-1}$ and $v_{south} = 76.4 \pm 0.3$ mas d$^{-1}$. The constant deceleration hypothesis gives initial velocities of $v_{north,0} = 35.9 \pm 0.5$ mas d$^{-1}$ and $v_{south,0} = 93.3 \pm 0.6$ mas d$^{-1}$. The acceleration values are $\dot{v}_{north,0} = -0.045 \pm 0.004$ mas d$^{-2}$ and $\dot{v}_{south,0} = -0.34 \pm 0.01$ mas d$^{-2}$.
Assuming a distance of 2.96 kpc \citep{Atri2020}, this corresponds to respective apparent initial velocities of 0.61~c and 1.59~c approximately.
The superluminal apparent velocity of the south jet, which is much higher than the apparent velocity for the north jet, suggests that the south jet is the approaching component of the ejection while the north jet is the receding component. The approaching / receding nature of the jets and their velocities are in accordance with what \citet{Bright2020} find, considering we have additional data at a later time to perform the fits.

To assess the goodness of fit of the linear and decelerated models to the observed data, we compute the chi-square statistic for both joint fits. 
The reduced chi-square is $\chi^2 _{\text{lin}}= 40$ for the linear model and $\chi^2 _{\text{dec}} = 7.3$ for the decelerated model. 
Even though both values are quite high, which can be attributed to the relatively small error bars of the \textit{Chandra} data points, the smaller value of $\chi^2 _{\text{dec}}$ suggests that the data is more in accordance with the constant deceleration hypothesis, following what was already hinted for the south jet. 
Moreover, the measure of a 14 mas angular separation between the two jets on MJD 58306.22 by \citet{Bright2020} implies an ejection date around MJD 58306.1, using the velocities obtained for both models. This is not compatible with the ejection date found for the ballistic motion, and strengthen the likeliness of a decelerated motion.

The addition of the \textit{Chandra} observations thus advocates strongly for the fact that the jets are decelerated, which could not be deduced by \citet{Bright2020} from the radio data alone. This implies the jets are probably emitted at the same time and then gradually slowed down, possibly by an interaction with their environment. 

The relatively high $\chi^2$ values could suggest that the deceleration is in fact not constant. Indeed, it has been suggested that X-ray binaries could be located in low-density bubbles \citep{Heinz2002}.  In that case, the deceleration would be enhanced when the jets interact with the denser ISM at the edge of the bubble. For instance, \citet{Wang2003}, \citet{Hao2009} and \citet{Steiner2012} found that the jets observed for \xte{} could be decelerated by interaction with the surrounding ISM, which accelerates the jet particles (similar findings were also invoked in \hh{}), possibly implying that a significant fraction of X-ray binaries could be surrounded by large-scale low-density cavities.

However, \citet{Bright2020} find that the decay rate of the radio emission coming from the jets is very slow, and attribute it to continuous on-going interaction with the ISM. This could advocate for a constant deceleration, especially as we have no data that could point towards a ballistic motion at the beginning of the propagation.

\subsection{Energetics} \label{4.2}

\begin{figure}[ht]%*
    \centering
    \includegraphics[trim={7.3cm 5.5cm 7.3cm 5.4cm}, clip, width=\linewidth]{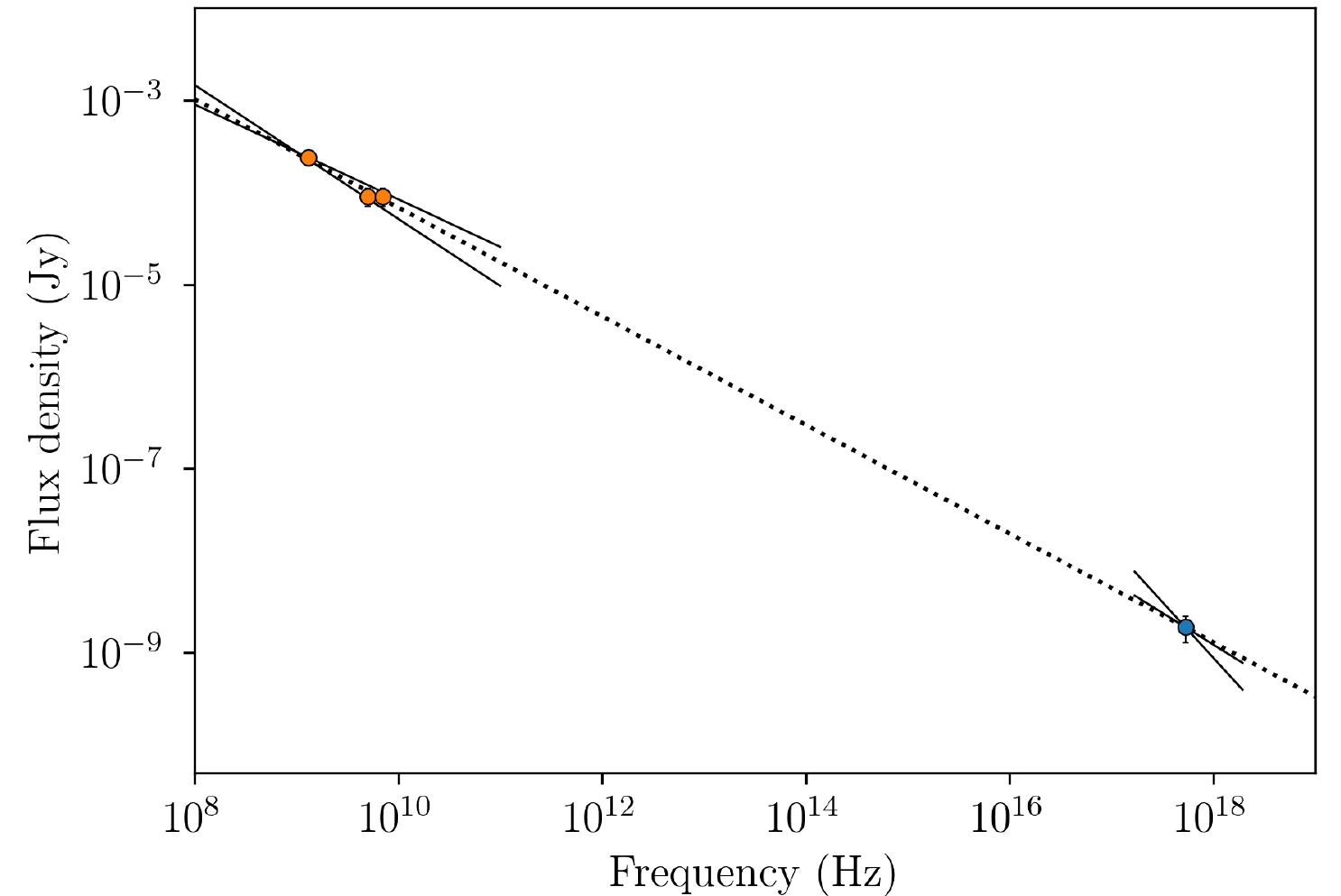}
 \caption{Radio to X-ray spectral energy distribution (SED) of the approaching jet in 2018 November. The dashed line represents the best fitting power-law model for the radio and X-ray data together. The full lines represent the power-law ranges coming from the errors on spectral indices for each wavelength domain. The flux densities correspond to the total emission of the jet observed at the three radio frequencies and at 2.2 keV for the X-rays.}
 \label{sed_jet}
\end{figure}

Using the available observations (\textit{Chandra}, VLA and MeerKAT), we are able to construct the broad-band spectra of the approaching and receding jets in 2018 November and of the receding jet in 2019 February. The three spectral energy distributions can be fitted with single power-laws with spectral indices of $\alpha = -0.59 \pm 0.01$ for the approaching jet in 2018 November, $\alpha =-0.65 \pm 0.02$ for the receding jet in 2018 November and $\alpha =-0.65 \pm 0.01$ for the receding jet in 2019 February. Figure \ref{sed_jet} displays the spectral energy distribution (SED) for the approaching jet in 2018 November. 

These spectral indices are consistent with what would be expected from optically thin synchrotron emission produced by shock-accelerated particles. Indeed, the apparent motion of the jets seems to favor a possible deceleration rather than a simple ballistic propagation. This could be due to the interaction of the jets with the ISM as observed previously in \xte{} \citep{Corbel2002, Migliori2017} and \hh{} \citep{Corbel2005}. This interaction would be responsible for the acceleration of the particles of the jets and thus for the observed broad-band synchrotron emission. 

To estimate the internal energy of the jets, we use the measured size obtained for the south jet using the \textit{Chandra} observation on 2018 November, see Figure~\ref{profil}.
The jet is then modeled by a truncated cone whose apex is at the central source, of $4.9"$ in height, $2.6"$ in width and with an opening angle of 7.1\degree. Using 2.96 kpc as the distance to \mx{} \citep{Atri2020}, this corresponds to a volume of $V = 1.4 \times 10^{51}$ cm$^3$, and a jet of $1.5 \times 10^{4}$ AU by $7.7 \times 10^{3}$ AU, which is in accordance with the estimation performed by \citet{Bright2020}. Their estimate was obtained 40 days before by inferring it from radio flux comparisons, while we obtained directly a physical size using \textit{Chandra} data. 

Using the slope of the spectral energy distribution (Figure \ref{sed_jet}) of the approaching jet which gives $\alpha = -0.59$, we estimate a total radiative luminosity of $L = 2.5 \times 10^{31}$ erg s$^{-1}$ between $\nu_1 =$ 1.3 GHz and $\nu_2 = 5.2 \times 10^{8}$ GHz.
Under standard hypothesis of equipartition and assuming all the energy is stored in the electrons (no energy carried by the protons), we can estimate the minimum internal energy of the synchrotron-emitting plasma.
Following \citet{Fender1999}, we estimate the parameters of the synchrotron emission with the formulae in \citet{Longair2011} (see section 16.5), using the opposite convention for the sign of $\alpha$. This yields a minimum internal energy of $5.1 \times 10^{41}$ erg and an equipartition magnetic field of the order of $2.0 \times 10^{-4}$ G. 
This implies radiating electrons with Lorentz factor of $\sim$  $3.1 \times 10^{7}$, i.e.  electrons accelerated up to energies > 10 TeV, and a cooling time scale of twenty-two years. This also leads to an estimate of $1.0 \times 10^{44}$ electrons in the jets, and if there is one proton per electron, we deduce a mass of the plasma of $\sim$  $1.7 \times 10^{20}$ g.
The total energy in the electrons would thus be $2.9 \times 10^{41}$ erg and the energy in the magnetic field $2.2 \times 10^{41}$ erg (consistent with the estimates from \citealt{Bright2020} using solely the radio observations). This strengthens their finding that the minimum internal energy of the jet is significantly ($\sim 10^4$ times) larger that the energy inferred from the radio flare believed to be the origin of the ejecta \citep{Bright2018}. 
Unless a significant fraction of the energy is radiated in a different wavelength during the launch (e.g. in X-rays, \citet{Homan2020} report a small flare in the 7-12 keV band just before the radio flare), this suggests the majority of the energy of the jets is not radiated and is released once they interact with the surrounding medium.
Furthermore, the above quantities are consistent with what was derived in \xte{} \citep{Tomsick2003} and \hh{} \citep{Corbel2005}, suggesting a common mechanism could be at play in the different sources displaying radio to X-ray jets.

\FloatBarrier

\acknowledgments

This research has made use of data obtained from the \textit{Chandra} X-ray Observatory (ObsId 20207, 20208, 22080, 21203, 21205), and software provided by the Chandra X-ray Center (CXC) in the application packages CIAO and Sherpa.
The MeerKAT telescope is operated by the South African Radio Astronomy Observatory, which is a facility of the National Research Foundation, an agency of the Department of Science and Innovation.
The National Radio Astronomy Observatory is a facility of the National Science Foundation operated under cooperative agreement by Associated Universities, Inc.
We acknowledge the use of the Nançay Data Center, hosted by the Nançay Radio Observatory (Observatoire de Paris-PSL, CNRS, Université d’Orléans), and also supported by Région Centre-Val de Loire. 
ME, SC and ET acknowledge financial support from the UnivEarthS Labex program of Université de Paris (ANR-10-LABX-0023 and ANR-11-IDEX-0005-02).
PK, EG and JH acknowledge financial support that was provided by the National Aeronautics and Space Administration through Chandra Award Numbers GO8-19033X, GO8-19027B and GO9-20027B issued by the Chandra X-ray Center, which is operated by the Smithsonian Astrophysical Observatory for and on behalf of the National Aeronautics Space Administration under contract NAS8-03060.
PGJ acknowledges funding from the European Research Council under ERC Consolidator Grant agreement no 647208.
JCAM-J is the recipient of an Australian Research Council Future Fellowship (FT140101082) funded by the Australian government.
We would like to thank the referee for a careful reading of the manuscript and for comments greatly improving its quality.

\facilities{CXO, VLA, MeerKAT}
\software{CIAO \citep{Fruscione2006}, Sherpa \citep{Freeman2001}, XSPEC \citep{Arnaud1996}, CASA \citep{McMullin2007}, APLpy \citep{Robitaille2012}, Astropy \citep{astropy:2013,astropy:2018}}

\clearpage

\bibliography{sources}
\bibliographystyle{aasjournal}

\end{document}